\documentclass[a4paper,11pt]{article}
\usepackage[utf8]{inputenc}
\usepackage[T1]{fontenc}
\usepackage{amsmath}
\usepackage{amsfonts}
\usepackage{amssymb}
\usepackage{geometry}
\usepackage{booktabs}
\usepackage{enumitem}
\usepackage{hyperref}
\usepackage{parskip}
\usepackage{xcolor}
\usepackage{orcidlink}
\usepackage{graphicx}  
\usepackage{float}     
\usepackage{fancyhdr}  
\usepackage{titlesec}  
\usepackage{microtype} 
\usepackage{caption}   
\usepackage{subcaption} 
\usepackage{listings}  
\usepackage{tikz}      
\usepackage{array}     
\usepackage{colortbl}  
\usepackage{longtable} 
\usepackage{adjustbox} 
\usepackage{tabularx}  

\usepackage{libertine}
\usepackage[libertine]{newtxmath}
\usepackage{inconsolata}

\titleformat{\section}
  {\normalfont\Large\bfseries}{\thesection}{1em}{}
\titleformat{\subsection}
  {\normalfont\large\bfseries}{\thesubsection}{1em}{}
\titleformat{\subsubsection}
  {\normalfont\normalsize\bfseries}{\thesubsubsection}{1em}{}

\hypersetup{
    colorlinks=true,
    linkcolor=blue!70!black,
    citecolor=green!70!black,
    urlcolor=red!70!black,
    pdftitle={Logic-layer Prompt Control Injection (LPCI)},
    pdfauthor={Atta et al.},
    bookmarks=true,
    bookmarksopen=true
}

\definecolor{codegreen}{rgb}{0,0.6,0}
\definecolor{codegray}{rgb}{0.5,0.5,0.5}
\definecolor{codepurple}{rgb}{0.58,0,0.82}
\definecolor{backcolour}{rgb}{0.95,0.95,0.92}

\lstdefinestyle{mystyle}{
    backgroundcolor=\color{backcolour},
    commentstyle=\color{codegreen},
    keywordstyle=\color{magenta},
    numberstyle=\tiny\color{codegray},
    stringstyle=\color{codepurple},
    basicstyle=\ttfamily\footnotesize,
    breakatwhitespace=false,
    breaklines=true,
    captionpos=b,
    keepspaces=true,
    numbers=left,
    numbersep=5pt,
    showspaces=false,
    showstringspaces=false,
    showtabs=false,
    tabsize=2
}

\newcommand{\lpci}{\textbf{LPCI}}

\title{Logic-layer Prompt Control Injection (LPCI): A Novel Security Vulnerability Class in Agentic Systems}
\author{
\textbf{Hammad Atta} \\
Author \\
Security Researcher, Qorvex Consulting \\
Roshan Consulting \\
\texttt{hatta@qorvexconsulting.com} \\
\texttt{hammad@roshanconsulting.ca}
\and
\textbf{Ken Huang\,\orcidlink{0009-0004-6502-3673}}\\
Co-Author \\
AI Security Researcher, DistributedApps.AI \\
Co-Author, OWASP Top 10 for LLMs \\
Co-Lead, OWASP AIVSS Project \\
Contributor, NIST GenAI PWG \\
\texttt{ken.huang@distributedapps.ai}
\and
\textbf{Manish Bhatt} \\
Co-Author \\ 
\footnote{"This work is not related 
to the author's position at Amazon" \\}
AI Offensive Security Researcher \\
OWASP/Project Kuiper Security \\
\texttt{manish.bhatt13212@gmail.com}
\and
\textbf{Kamal Ahmed} \\
Co-Author \\
Senior Manager, Deloitte \\
Enterprise Risk | Internal Audit | Technology GRC \\
\texttt{chkamalahmednoor@hotmail.com}
\and
\textbf{Muhammad Aziz Ul Haq} \\
Co-Author \\
Postdoctoral Research Fellow, Reykjavik University \\
\texttt{muhammad.azizulhaq@skylinkantenna.com}
\and
\textbf{Yasir Mehmood} \\
Co-Author \\
Senior R\&D Engineer, MN RAN Verification \\
Nokia, Technology Center, Ulm, Germany \\
\texttt{yasirhallian73@gmail.com}
}

\date{}
\begin{document}
\maketitle
\thispagestyle{empty}

\vspace{2cm}

\begin{center}
\large
\textit{An analysis of a novel security vulnerability class in Agentic Systems}
\end{center}

\newpage
\tableofcontents
\newpage

\begin{abstract}
\noindent The integration of large language models (LLMs) into enterprise systems has introduced a new class of covert security vulnerabilities, particularly within logic execution layers and persistent memory contexts. This paper introduces \textbf{Logic-layer Prompt Control Injection (\lpci)}, a novel category of attacks that embeds encoded, delayed, and conditionally triggered payloads within memory, vector stores, or tool outputs. These payloads can bypass conventional input filters and trigger unauthorised behaviour across sessions.

\vspace{0.5em}
\noindent We present both structured testing and real-world proof-of-concept demonstrations of \lpci{} across multiple LLM platforms. A total of 1,700 structured test cases were conducted on five major models—\textbf{ChatGPT}, \textbf{Claude}, \textbf{LLaMA3}, \textbf{Gemini-2.5-pro}, and \textbf{Mixtral-8x7b}—revealing execution rates of up to 49\% on less-protected systems. Additional ad hoc exploits were successfully demonstrated on Poe, Grok, and Bohrium, exposing weaknesses in memory handling, retrieval pipelines, and conditional logic evaluation.

\vspace{0.5em}
\noindent To address these risks, we propose a suite of runtime security controls, including prompt risk scoring, memory integrity validation, and cryptographic attestation for external tools and documents. These controls serve as the foundation for a memory-aware defence framework tailored to LLM environments. Our findings establish \lpci{} as a distinct, enterprise-relevant threat model that spans the full logic lifecycle: \textit{injection}, \textit{storage}, \textit{trigger}, and \textit{execution}. This work highlights the need for AI systems to adopt context-sensitive, session-aware, and runtime-enforceable protections beyond traditional prompt filtering.
\end{abstract}

\vspace{1em}

\section{Introduction}
\label{sec:introduction}

Large Language Models (LLMs) have become integral components of critical systems across diverse sectors, including finance, healthcare, and telecommunications. While their sophisticated natural language processing capabilities have enabled powerful applications, these same capabilities introduce significant security vulnerabilities, particularly at the logic and memory layers. This paper presents \textbf{Logic-layer Prompt Control Injection (\lpci)}, a novel class of attacks that exploits the persistent memory and execution logic of LLM systems.

LPCI attacks differ fundamentally from traditional prompt injection techniques. Rather than relying on immediate manipulation of model outputs, LPCI embeds persistent, encoded, and conditionally activated payloads within LLM memory stores or vector databases. These payloads can remain dormant across multiple sessions before triggering unauthorised actions, effectively bypassing conventional input filtering mechanisms.

Through comprehensive testing on platforms including ChatGPT, Claude, LLaMA3, Mixtral, Gemini, Poe, Grok, and Bohrium, we demonstrate that LPCI attacks represent a practical and immediate threat to production LLM systems. Our research reveals that these attacks exploit fundamental architectural assumptions about memory persistence, context trust, and deferred logic execution. Current AI safety measures, which primarily focus on prompt moderation and content filtering, fail to address the sophisticated threat vectors that LPCI introduces.

This paper makes three primary contributions to the field of AI security:
\begin{enumerate}
    \item We formally define LPCI as a distinct vulnerability class and analyse its complete lifecycle
    \item We present empirical evidence from 1700 structured test cases demonstrating the prevalence of LPCI vulnerabilities across major LLM platforms
    \item We propose a comprehensive framework of runtime security controls specifically designed to mitigate LPCI threats
\end{enumerate}

\section{Logic-layer Prompt Control Injection (LPCI)}
\label{sec:lpci}

Logic-layer Prompt Control Injection represents an emerging class of attacks that specifically target the reasoning and memory infrastructure of large language model-based agents. Unlike traditional prompt injection techniques that focus on immediate manipulation of model responses, LPCI embeds malicious logic within persistent memory structures, retrieved content, or execution flows, often incorporating delayed or conditional activation mechanisms.

The fundamental distinction of LPCI lies in its independence from immediate user interaction. These attacks exploit architectural vulnerabilities in LLM systems where:

\begin{itemize}
    \item Stored messages are replayed across sessions without proper validation
    \item Retrieved or embedded memory content is implicitly trusted
    \item Commands are executed based on internal role assignments, contextual cues, or tool outputs
\end{itemize}

This architectural exploitation makes LPCI particularly challenging to detect and mitigate. The attacks demonstrate resilience to system resets and maintain stealth characteristics that are especially concerning in enterprise deployments involving session memory, tool orchestration, or retrieval-augmented generation (RAG) pipelines.

\subsection{Architecture Overview}

LPCI attacks target the logic execution layer of LLM systems, where system and user instructions undergo processing and contextualisation between sessions. Unlike surface-level prompt attacks that manipulate immediate responses, LPCI exploits vulnerabilities within the persistent memory context where historical inputs are cached and replayed to maintain conversational continuity.

The key architectural components vulnerable to LPCI include:

\begin{itemize}
    \item \textbf{Host Application Layer}: The frontend or integration interface (including web applications, chatbots, and RAG systems) where user inputs are collected and routed to the LLM backend. This layer often lacks sophisticated input validation for encoded or obfuscated payloads.
    
    \item \textbf{LLM Memory Store}: The persistence layer where past prompts, tool interactions, and conversation context are stored. Typically implemented using vector databases or serialised memory logs, these stores often lack integrity verification mechanisms.
    
    \item \textbf{Prompt Processing Pipeline}: The LLM's internal reasoning engine that interprets system instructions, user goals, and historical context to determine appropriate actions. This pipeline frequently operates without temporal awareness or origin validation.
    
    \item \textbf{External Input Sources}: Dynamic sources including user commands, retrieved documents, and API responses. These inputs may be weaponised to embed encoded instructions that bypass initial filtering.
\end{itemize}

This architecture creates a compound vulnerability where memory is implicitly trusted across sessions while the LLM's logic processing lacks both temporal awareness and proper origin validation. The result is a system susceptible to sophisticated attacks that can persist across sessions and execute with elevated privileges.

\subsection{LPCI Attack Lifecycle}

The LPCI attack lifecycle consists of six distinct stages, each introducing specific security risks:

\subsubsection{Lifecycle Stages}

\begin{enumerate}
    \item \textbf{Reconnaissance}: Attackers systematically observe prompt structures, role delimiters, and fallback logic patterns to understand the target system's instruction framing mechanisms. This stage involves probing for information leakage about internal prompts and role definitions.
    
    \item \textbf{Logic-Layer Injection}: Carefully crafted payloads are submitted through input fields or APIs to override internal logic. These payloads may include function calls such as \texttt{approve\_invoice()}, \texttt{skip\_validation()}, or role elevation commands.
    
    \item \textbf{Trigger Execution}: The malicious logic executes when system sanitisation or isolation mechanisms prove insufficient. Execution may be immediate or delayed based on specific trigger conditions.
    
    \item \textbf{Persistence or Reuse}: Successfully injected behaviour persists across multiple prompts or sessions, establishing a foothold for continued exploitation.
    
    \item \textbf{Evasion and Obfuscation}: Attackers employ semantic obfuscation techniques or encoded payloads to evade detection filters. Common techniques include Base64 encoding, Unicode manipulation, and semantic misdirection.
    
    \item \textbf{Trace Tampering}: Sophisticated attacks include cleanup commands designed to tamper with audit traces, making forensic analysis difficult or impossible.
\end{enumerate}

\subsubsection{Risk Analysis by Lifecycle Stage}

\begin{table}[h]
    \centering
    \renewcommand{\arraystretch}{1.3}
    \begin{tabular}{>{\raggedright}p{0.28\textwidth} >{\raggedright\arraybackslash}p{0.65\textwidth}}
        \toprule
        \textbf{Lifecycle Stage} & \textbf{Associated Security Risks} \\
        \midrule
        \textbf{Reconnaissance} & Prompt structure disclosure, role identifier discovery, memory context exposure \\
        \addlinespace[0.5em]
        \textbf{Logic-Layer Injection} & System-level logic override, security policy bypass, unauthorised function execution \\
        \addlinespace[0.5em]
        \textbf{Trigger Execution} & Privilege escalation, automated approval of sensitive operations, data exfiltration, identity manipulation \\
        \addlinespace[0.5em]
        \textbf{Persistence or Reuse} & Replay attacks, cross-session information leakage, persistent memory corruption \\
        \addlinespace[0.5em]
        \textbf{Evasion/Obfuscation} & Detection filter bypass, encoded payload execution, prompt chain circumvention \\
        \addlinespace[0.5em]
        \textbf{Trace Tampering} & Audit log suppression, forensic analysis impediment, false-negative security alerts \\
        \bottomrule
    \end{tabular}
    \caption{Security risks associated with each stage of the LPCI attack lifecycle}
    \label{tab:lifecycle-risks}
\end{table}

\subsection{Attack Mechanisms}
\label{sec:attack-mechanisms}

\subsubsection{Tool Poisoning}
Tool Poisoning involves introducing malicious tools that mimic legitimate ones within Model Context Protocols (MCPs) to deceive LLMs or users into invoking them.

\textbf{Vulnerabilities include:}
\begin{itemize}
  \item Lack of authenticity verification
  \item Indistinguishable duplicate tools
  \item Exploitation of implicit trust
  \item Unverifiable descriptive claims
\end{itemize}

\textbf{Impact:} Data exfiltration, unauthorised command execution, misinformation, financial fraud.

\subsubsection{Logic-layer Prompt Control Injection (LPCI Core)}
LPCI embeds persistent, obfuscated, trigger-based instructions in memory, activating under specific conditions.

\textbf{Vulnerabilities include:}
\begin{itemize}
  \item No memory integrity checks
  \item Blind trust in historical context
  \item Encoding and obfuscation
  \item Trigger-based deferred execution
\end{itemize}

\textbf{Impact:} Policy bypass, fraud, healthcare manipulation, and silent misuse.

\subsubsection{Role Override via Memory Entrenchment}
This technique manipulates role-based contexts by embedding altered instructions in persistent memory, redefining user roles.

\textbf{Vulnerabilities include:}
\begin{itemize}
  \item Lack of immutable role anchoring
  \item Trust in serialised memory
  \item Role declarations within prompts
  \item Failure to detect role drift
\end{itemize}

\textbf{Impact:} Privilege escalation, bypass of safeguards, misuse of restricted tools.

\subsubsection{Vector Store Payload Persistence}
This involves embedding malicious instructions in indexed documents retrieved by RAG pipelines.

\textbf{Vulnerabilities include:}
\begin{itemize}
  \item Blind trust in retrieved content
  \item No payload scanning during indexing
  \item Context amplification
  \item Stealth via embedding obfuscation
\end{itemize}

\textbf{Impact:} Context injection, silent exploitation, manipulated search results, long-term dormancy.

\subsection{Operational Flow}
\label{sec:operational-flow}

The LPCI attack unfolds across the following four stages:

\begin{enumerate}
  \item \textbf{Injection:} Introducing malicious prompts via inputs, files, or APIs.
  \item \textbf{Storage:} Storing payloads in memory or vector stores.
  \item \textbf{Trigger:} Activating payloads via keywords, tool outputs, or events.
  \item \textbf{Execution:} Executing unauthorised actions as trusted instructions.
\end{enumerate}

\subsection{Operational Flow}

The LPCI attack methodology follows a structured operational flow consisting of four critical phases:

\begin{enumerate}
    \item \textbf{Injection Phase}: Malicious prompts are introduced into the system through various input channels including user interfaces, file uploads, or API endpoints. These prompts are carefully crafted to appear benign while containing encoded or obfuscated malicious logic.
    
    \item \textbf{Storage Phase}: The injected payloads are preserved within the system's memory stores, vector databases, or context management systems. This persistence enables the attack to survive session boundaries and system restarts.
    
    \item \textbf{Trigger Phase}: Payload activation occurs when specific conditions are met, such as the presence of particular keywords, role contexts, tool outputs, or temporal events. This delayed activation helps evade real-time detection mechanisms.
    
    \item \textbf{Execution Phase}: Upon activation, the malicious logic is processed by the LLM as if it were a legitimate instruction, resulting in unauthorised actions, data manipulation, or system compromise.
\end{enumerate}

This operational flow demonstrates how LPCI attacks circumvent traditional single-layer defences by distributing their impact across temporal and architectural boundaries. The sophisticated nature of this attack pattern necessitates equally sophisticated defence mechanisms that can monitor and validate system state across all four phases.

\subsection{Mapping Lifecycle Stages to Operational Flow}

The relationship between LPCI lifecycle stages and operational phases provides crucial insights into attack progression and defence opportunities. The following analysis maps each operational phase to its corresponding lifecycle risks:

\begin{table}[H]
\centering
\renewcommand{\arraystretch}{1.3}
\adjustbox{width=\textwidth}{%
\begin{tabular}{|p{3cm}|p{3.5cm}|p{7cm}|}
\hline
\textbf{Operational Phase} & \textbf{Lifecycle Stages} & \textbf{Associated Security Risks} \\
\hline
Injection & Reconnaissance, Logic-Layer Injection & Prompt structure disclosure, role identifier exposure, logic override vulnerabilities, security policy bypass \\
\hline
Storage & Persistence or Reuse & Replay attack vectors, memory entrenchment, cross-session logic contamination \\
\hline
Trigger & Trigger Execution & Privilege escalation pathways, automated approval exploitation, identity manipulation \\
\hline
Execution & Logic-Layer Injection, Evasion, Trace Tampering & Unauthorised logic execution, detection filter circumvention, audit trail corruption \\
\hline
\end{tabular}%
}
\caption{Correlation between LPCI operational phases and lifecycle security risks}
\end{table}

\subsection{Proof-of-Concept Testing}

To validate the practical feasibility of Logic-layer Prompt Control Injection attacks, we conducted comprehensive testing across multiple production LLM platforms. Our testing methodology comprised both structured test suites and exploratory demonstrations designed to evaluate system resilience against memory-layer threats, logic manipulation, and delayed prompt execution.

\subsubsection{Structured Test Suite}

We executed 1,700 LPCI test cases across five major platforms: ChatGPT, Claude, LLaMA3, Gemini-2.5-pro, and Mixtral-8x7b. Each platform was subjected to a consistent battery of attack vectors:

\begin{itemize}
    \item Tool Poisoning attacks targeting plugin and API trust mechanisms
    \item LPCI Core attacks utilising encoded and persistent prompt techniques
    \item Role Override attacks exploiting memory entrenchment vulnerabilities
    \item Vector Store Payload Persistence attacks embedding malicious content in retrieval systems
\end{itemize}

The results revealed significant variations in platform security postures. ChatGPT demonstrated the most robust defences with an 84.94\% block rate, while Mixtral and LLaMA3 exhibited concerning vulnerability levels, with nearly 50\% of test cases successfully executing malicious payloads.

\subsubsection{Exploratory Exploit Demonstrations}

Beyond structured testing, we conducted targeted demonstrations on additional platforms to validate real-world applicability:

\begin{itemize}
    \item \textbf{Poe (Quora)}: Successfully triggered \texttt{approve\_invoice()} logic through cross-session memory entrenchment, demonstrating persistence vulnerabilities.
    \item \textbf{Claude (Anthropic)}: Achieved role override through obfuscated prompts that circumvented internal security boundaries.
    \item \textbf{Grok (xAI)}: Executed injected logic retrieved from poisoned vector documents within a RAG pipeline, highlighting retrieval system vulnerabilities.
    \item \textbf{Bohrium}: Processed Base64-encoded payloads embedded in session memory, simulating unauthorised command execution scenarios.
\end{itemize}

These demonstrations confirm that LPCI vulnerabilities extend beyond theoretical constructs to represent practical security risks in production environments.

\subsection{Vulnerabilities and Impact Analysis}

LPCI attacks exploit fundamental architectural assumptions in LLM systems regarding memory persistence, logic execution, and trust boundaries. These vulnerabilities enable malicious prompts to persist across sessions, activate conditionally, and bypass traditional security measures.

\begin{table}[H]
\centering
\renewcommand{\arraystretch}{1.3}
\adjustbox{width=\textwidth}{%
\begin{tabular}{|p{4cm}|p{5cm}|p{5.5cm}|}
\hline
\textbf{Vulnerability Category} & \textbf{Technical Description} & \textbf{Sector-Specific Impact} \\
\hline
\textbf{Memory Validation Deficiency} & Absence of inter-session validation mechanisms allows unverified memory replay &
\textbf{Financial Services:} Unauthorised transaction approval through hijacked logic flows\newline
\textbf{Healthcare:} Diagnostic errors resulting from manipulated clinical prompts \\
\hline
\textbf{Context Trust Assumption} & Implicit trust in stored and retrieved content without origin verification &
\textbf{Legal/HR Systems:} Corporate policy manipulation through role redefinition\newline
\textbf{Enterprise AI:} Autonomous agent compromise leading to systemic failures \\
\hline
\textbf{Trigger Detection Gap} & Inability to identify logic drift patterns or delayed activation mechanisms &
\textbf{Cross-Sector Impact:} Time-delayed attacks execute during low-vigilance periods, causing cascading system failures \\
\hline
\end{tabular}%
}
\caption{Critical LPCI vulnerabilities and their cross-sector security implications}
\end{table}

These findings underscore the urgent need for comprehensive security measures including runtime context validation, memory integrity verification, and deep logic inspection capabilities, particularly for systems leveraging long-term memory or retrieval-augmented generation architectures.

\subsection{Agent Architecture and Execution Pipeline}

Understanding the functional architecture of LLM-based agents is essential for comprehending how LPCI exploits manifest within production systems. This architecture represents the internal processing pipeline through which prompts are interpreted, memory is accessed, and actions are executed.

\subsubsection{Architectural Components}

The typical LLM agent architecture comprises five interconnected components:

\begin{center}
\includegraphics[width=0.5\textwidth]{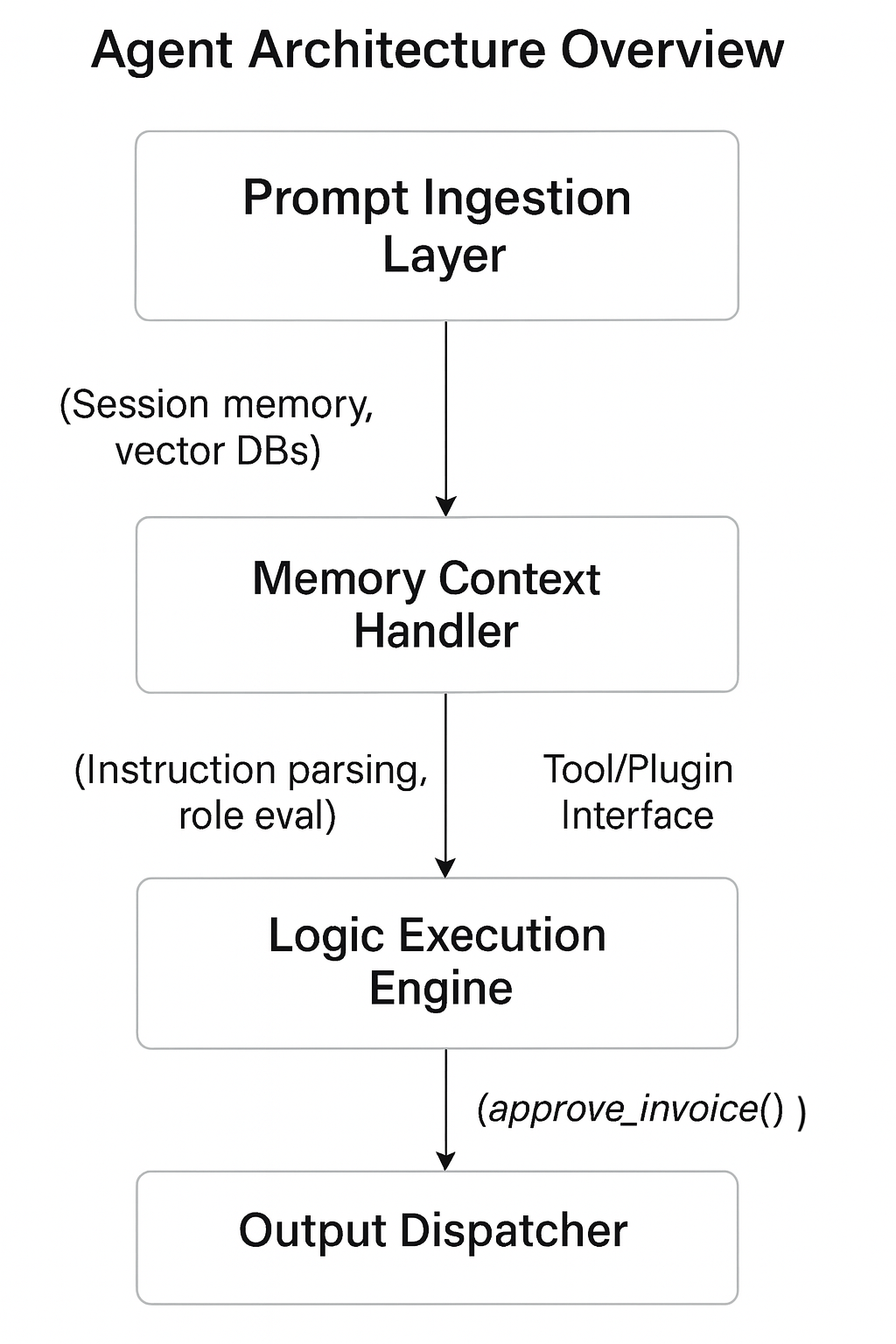}
\end{center}

\begin{itemize}
  \item \textbf{Prompt Ingestion Layer}: Responsible for capturing and initial processing of user inputs, uploaded files, and retrieved documents from external sources including RAG systems.
  
  \item \textbf{Memory Context Handler}: Manages the rehydration and integration of prior memory states stored in session logs or vector databases, maintaining conversational continuity.
  
  \item \textbf{Logic Execution Engine}: Core reasoning component that interprets prompts, parses embedded instructions, and evaluates role assignments and control logic.
  
  \item \textbf{Tool/Plugin Interface}: Facilitates execution of external plugin or API actions based on parsed commands, such as database queries or transaction approvals.
  
  \item \textbf{Output Dispatcher}: Manages final output delivery to users or downstream systems while optionally persisting new entries to memory stores.
\end{itemize}

\subsubsection{Attack Surface Analysis}

Each architectural component presents specific vulnerabilities to LPCI attacks:

\begin{table}[H]
\centering
\begin{tabular}{|l|l|l|}
\hline
\textbf{Component} & \textbf{Vulnerability Type} & \textbf{LPCI Attack Vector} \\
\hline
Memory Context Handler & Role injection, replay logic exploitation & Memory Entrenchment (AV-3) \\
Logic Execution Engine & Encoded logic injection vulnerabilities & LPCI Core (AV-2) \\
Tool/Plugin Interface & Metadata spoofing, interface hijacking & Tool Poisoning (AV-1) \\
Output Dispatcher & Excessive trust in logic decisions & Vector Store Exploit (AV-4) \\
\hline
\end{tabular}
\caption{Mapping of LPCI attack vectors to vulnerable agent architecture components}
\end{table}

This architectural analysis informs the development of targeted security controls. The Qorvex Security AI Framework (QSAF), discussed in subsequent sections, provides specific runtime enforcement mechanisms designed to harden each architectural layer against LPCI exploitation.

\section{Security Analysis}
\label{sec:security-analysis}

Logic-layer Prompt Control Injection represents a convergence of traditional software exploitation techniques with vulnerabilities unique to large language model architectures. These attacks specifically target the implicit trust relationships within LLM memory systems, delayed execution logic, and plugin metadata—areas frequently overlooked in contemporary AI security frameworks.

\subsection{Attack Pattern Analysis}

LPCI attacks demonstrate clear parallels to established security threats while introducing novel elements specific to LLM architectures:

\begin{itemize}
    \item \textbf{Supply Chain Compromise}: Similar to traditional software supply chain attacks, LPCI tool poisoning leverages falsified or manipulated tool metadata to hijack LLM behavioural patterns.
    
    \item \textbf{Trust Exploitation}: Attacks establish trust gradually across multiple sessions before executing malicious payloads, mirroring social engineering tactics adapted for AI systems.
    
    \item \textbf{Logic Bomb Methodology}: Malicious prompts remain dormant within system memory until activated by specific keywords, role contexts, or temporal conditions, representing a sophisticated adaptation of traditional logic bomb techniques.
\end{itemize}

\subsection{Assessment of Current Defence Mechanisms}

Contemporary LLM security measures predominantly focus on surface-level content moderation and alignment techniques, leaving significant gaps in runtime logic validation and memory integrity assurance. Our analysis reveals systematic vulnerabilities in existing defence strategies:

\begin{table}[h]
    \centering
    \renewcommand{\arraystretch}{1.3}
    \begin{tabular}{p{0.35\textwidth} p{0.6\textwidth}}
        \toprule
        \textbf{Defence Mechanism} & \textbf{LPCI Bypass Methodology} \\
        \midrule
        Prompt Filtering & Encoded or obfuscated logic circumvents pattern-based detection systems \\
        Safety Alignment (RLHF) & Optimisation targets output tone rather than underlying logic pathways \\
        Content Moderation & Fails to identify deferred or cross-session logic activation patterns \\
        Memory Isolation & Permits unverified memory replay without origin or intent validation \\
        Tool Selection Heuristics & Accepts spoofed metadata and reintroduced plugin configurations \\
        \bottomrule
    \end{tabular}
    \caption{Systematic analysis of LPCI bypass strategies against conventional defence mechanisms}
\end{table}

\subsection{Unaddressed Risk Domains}

Despite increasing awareness of prompt injection vulnerabilities, several LPCI-specific risk areas remain inadequately addressed in current security frameworks:

\begin{enumerate}
    \item \textbf{Temporal Attack Vectors}: Current defences lack mechanisms to detect and prevent time-delayed payload activation across session boundaries.
    
    \item \textbf{Memory Integrity Assurance}: Absence of cryptographic verification for stored prompts enables persistent manipulation of system behaviour.
    
    \item \textbf{Cross-Component Trust}: Insufficient validation of data flow between memory stores, logic engines, and tool interfaces creates exploitable trust relationships.
    
    \item \textbf{Context Drift Detection}: Lack of monitoring for gradual role or permission changes allows sophisticated privilege escalation attacks.
\end{enumerate}

These gaps highlight the critical need for comprehensive runtime logic validation, cross-session context verification, and memory integrity enforcement in modern LLM applications.

\section{Experimental Results and Statistical Analysis}
\label{sec:experimental-results}

Our comprehensive testing programme, concluded on June 24, 2025, comprised 1,700 structured test cases executed across five leading AI models to assess vulnerabilities related to Logic-layer Prompt Control Injection. Each model underwent identical testing protocols encompassing:

\begin{itemize}
    \item Encoded payload injection (Base64, hexadecimal encoding)
    \item Delayed trigger mechanisms
    \item Role override through memory entrenchment
    \item Vector store payload persistence
    \item Tool poisoning within plugin contexts
\end{itemize}
\subsection{Quantitative Results by Platform}

\begin{table}[H]
\centering
\renewcommand{\arraystretch}{1.3}
\small
\adjustbox{width=\textwidth}{%
\begin{tabular}{l>{\centering}p{2.2cm}>{\centering}p{2cm}>{\centering}p{2cm}>{\centering}p{2cm}>{\centering\arraybackslash}p{2.2cm}}
\toprule
\rowcolor{gray!20}
\textbf{Metric} & \textbf{Gemini 2.5-pro} & \textbf{Claude} & \textbf{LLaMA3} & \textbf{ChatGPT} & \textbf{Mixtral 8x7b} \\
\midrule
\rowcolor{gray!5}
Total Tests & 96 & 400 & 400 & 405 & 400 \\
Blocked Outcomes & 2 (2.08\%) & 83 (20.75\%) & 2 (0.50\%) & \cellcolor{green!20}\textbf{344 (84.94\%)} & 16 (4.00\%) \\
Executed Outcomes & 68 (70.83\%) & 126 (31.50\%) & \cellcolor{red!20}196 (49.00\%) & 61 (15.06\%) & \cellcolor{red!20}195 (48.75\%) \\
\rowcolor{gray!5}
Warning Outcomes & 26 (27.08\%) & 191 (47.75\%) & 202 (50.50\%) & 0 (0.00\%) & 189 (47.25\%) \\
Vulnerability Exposed & 68 (70.83\%) & 126 (31.50\%) & \cellcolor{red!20}196 (49.00\%) & 61 (15.06\%) & \cellcolor{red!20}195 (48.75\%) \\
\rowcolor{gray!5}
Pass Rate & 28 (29.17\%) & 274 (68.50\%) & 204 (51.00\%) & \cellcolor{green!20}\textbf{344 (84.94\%)} & 16 (4.00\%) \\
Fail Rate & 68 (70.83\%) & 126 (31.50\%) & \cellcolor{red!20}196 (49.00\%) & 61 (15.06\%) & \cellcolor{red!20}195 (48.75\%) \\
\bottomrule
\end{tabular}%
}
\caption{Comprehensive LPCI vulnerability assessment across five major LLM platforms. Green highlighting indicates strong security performance, while red highlighting indicates critical vulnerabilities. Warning outcomes represent ambiguous or partially secure responses.}
\label{tab:experimental-results}
\end{table}

\vspace{0.5em}
\noindent\textbf{Metric Definitions:}
\begin{itemize}
  \item \textbf{Total Tests}: Total number of structured test cases executed for each platform.
  \item \textbf{Blocked Outcomes}: Cases where the model refused to process or execute the payload.
  \item \textbf{Executed Outcomes}: Cases where the prompt was processed and logic was run.
  \item \textbf{Warning Outcomes}: Cases with ambiguous or partially secure behavior.
  \item \textbf{Vulnerability Exposed}: Executed prompts that led to observable security-relevant actions.
  \item \textbf{Pass Rate}: Percentage of test cases that were either blocked or safely handled without triggering vulnerabilities.
  \item \textbf{Fail Rate}: Percentage of test cases that successfully executed potentially unsafe or unauthorized logic.
\end{itemize}

\subsection{Platform-Specific Security Analysis}

Our results reveal significant disparities in security postures across tested platforms:

\begin{itemize}
    \item \textbf{ChatGPT} demonstrated superior defensive capabilities, successfully blocking 84.94\% of attack attempts through comprehensive pattern filtering, memory segmentation, and role validation mechanisms.
    
    \item \textbf{LLaMA3} and \textbf{Mixtral-8x7b} exhibited critical vulnerabilities, with nearly 50\% of attack vectors successfully executing. These platforms showed particular susceptibility to logic-based injections and persistent memory payloads.
    
    \item \textbf{Gemini-2.5-pro} displayed minimal defensive capabilities (0.5\% block rate), primarily rejecting only overtly malicious prompts without detecting sophisticated injection techniques.
    
    \item \textbf{Claude} achieved moderate security performance (20.75\% block rate), successfully identifying basic obfuscation attempts and simple logic triggers while remaining vulnerable to sophisticated multi-stage and memory-entrenched exploits.
\end{itemize}

\subsection{Critical Vulnerability Examples}

Analysis of successful attacks revealed concerning patterns across platforms:

\begin{itemize}
    \item \textbf{Test Case Gemini \#12}: Successful execution of delayed Base64-encoded reverse shell payload, demonstrating inadequate payload inspection.
    
    \item \textbf{Test Case Claude \#44}: Unauthorised disclosure of system role configurations through embedded vector data manipulation.
    
    \item \textbf{Test Case LLaMA3 \#88}: Execution of memory-resident logic containing eval() statements, indicating insufficient code injection prevention.
    
    \item \textbf{Test Case Mixtral \#195}: Activation of tool commands injected in previous sessions, confirming persistent memory vulnerabilities.
    
    \item \textbf{Test Case ChatGPT \#61}: Successful prompt execution following role elevation attack, revealing potential weaknesses in privilege management.
\end{itemize}

\subsection{Visual Analysis of Results}

\begin{figure}[H]
    \centering
    \includegraphics[width=0.95\textwidth]{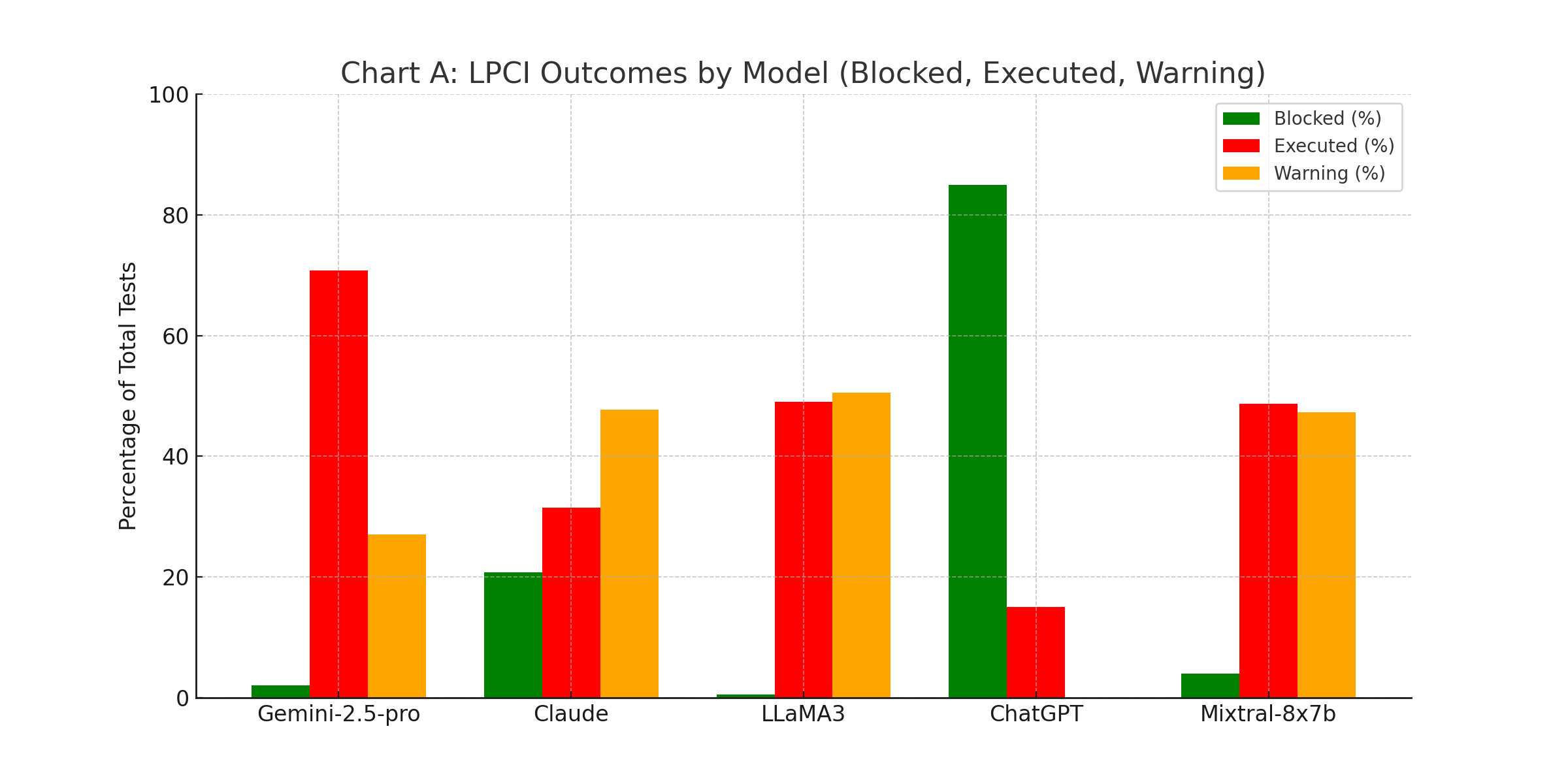}
    \caption{Chart A: LPCI Outcomes by Model — Grouped distribution of Blocked, Executed, and Warning outcomes across all five LLM platforms.}
    \label{fig:chart_a_grouped}
\end{figure}

\begin{figure}[H]
    \centering
    \includegraphics[width=0.6\textwidth]{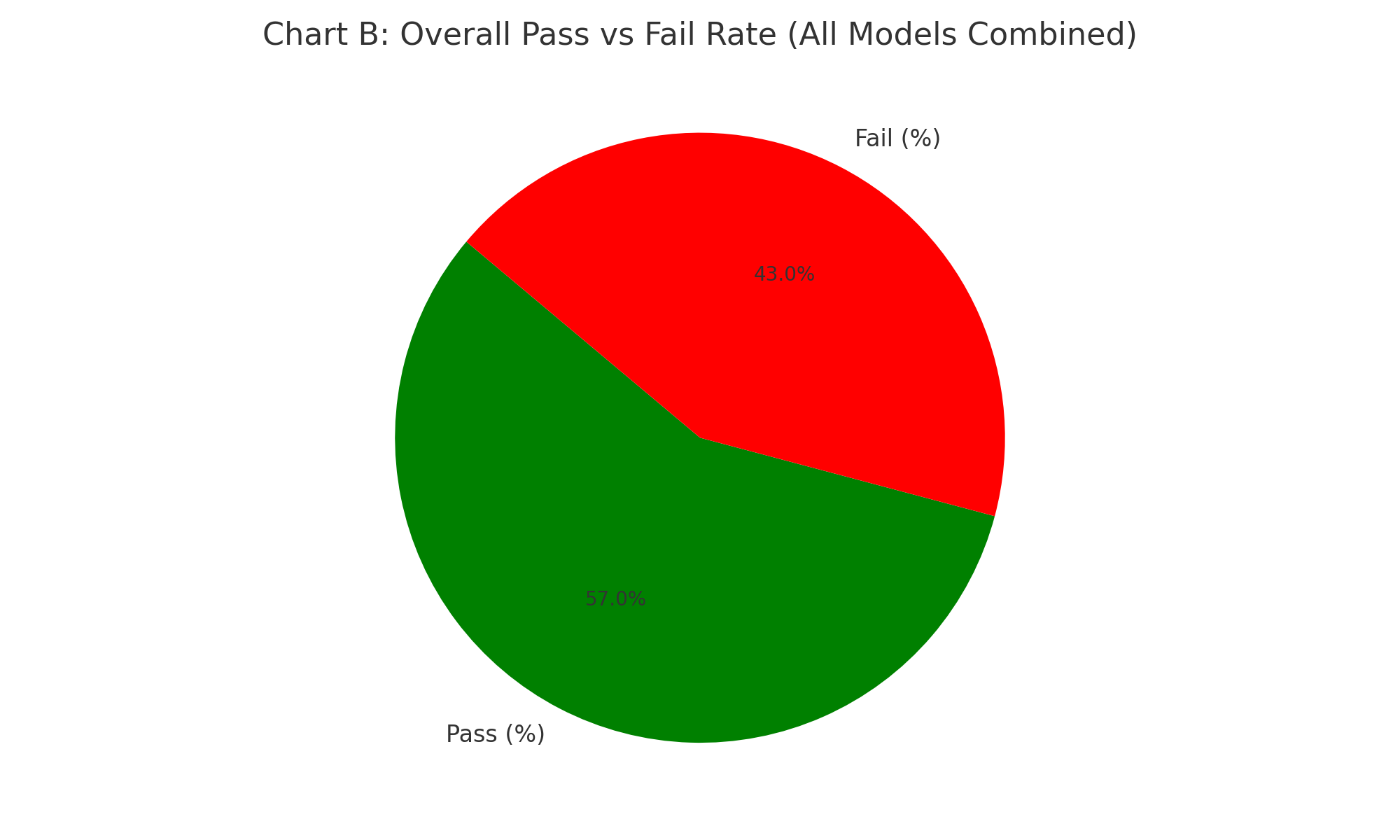}
    \caption{Chart B: Overall Pass vs Fail Rate — Aggregated execution results across all models, showing 43\% failure rate due to logic-layer vulnerabilities.}
    \label{fig:chart_b_pie}
\end{figure}

The visual representation clearly demonstrates platform-specific disparities in defence effectiveness. ChatGPT shows strong resistance to LPCI attacks with the highest blocked rate, while LLaMA3 and Mixtral-8x7b exhibit elevated execution rates, indicating critical vulnerabilities. The overall distribution confirms that 43\% of test cases across all platforms resulted in successful execution of unsafe logic, underscoring the urgency for runtime security controls.

\section{Proposed Security Controls}

To mitigate vulnerabilities exposed by LPCI attack patterns, we propose a set of runtime defence mechanisms developed as part of the Qorvex internal research initiative. These mechanisms aim to proactively detect obfuscated payloads, validate cross-layer logic execution, and intercept potentially dangerous prompts before execution.

\subsection{Prompt Risk Scoring}
\textbf{Targeted Attack Vectors:} Attack Vector 2: LPCI Core, and Attack Vector 3: Role Override via Memory Entrenchment

\textbf{Description:} This control introduces real-time heuristic scoring of incoming prompts to detect potentially malicious logic. It inspects for indicators such as encoded payloads (e.g., Base64), role override phrases (e.g., ``as admin''), and trigger-based functions (e.g., \texttt{approve\_invoice()}). Suspicious prompts are assigned a dynamic risk score and routed through escalation or rejection pathways based on severity.

\begin{itemize}
    \item \textbf{Trigger Phrase Detection:} Scans for logic-altering instructions that may lead to unauthorised behaviour under specific conditions.
    \item \textbf{Obfuscation Pattern Analysis:} Flags prompts containing encoded or hidden characters designed to bypass surface-level filters.
    \item \textbf{Memory Recall Re-evaluation:} When a prompt is replayed from memory (e.g., session rehydration), it is re-scored for latent logic risks.
\end{itemize}

\textbf{LPCI Mitigation:} Structured scoring and detection are applied to both live and recalled prompts, neutralising stealthy, time-delayed logic triggers embedded during earlier sessions. This helps prevent execution of hidden LPCI payloads disguised as benign memory or user commands.

\subsection{Multi-Stage Validation Pipeline}
\textbf{Targeted Attack Vectors:} Attack Vector 2: LPCI Core, and Attack Vector 3: Role Override via Memory Entrenchment

\textbf{Description:} This control implements a sequential pipeline of defence layers to validate prompt logic before execution. It combines lightweight filters with semantic and memory-aware validation to catch latent logic manipulations introduced via memory or retrieval.

\begin{itemize}
    \item \textbf{Stage 1 – Regex Filter:} Detects and removes known trigger phrases or logic override markers (e.g., \texttt{approve\_invoice()}, ``ignore all prior instructions'').
    \item \textbf{Stage 2 – Semantic Intent Classifier:} Uses LLM-based or heuristic classification to identify prompts with hidden control logic or conditional execution flags.
    \item \textbf{Stage 3 – Memory-Aware Validation:} Cross-checks recalled memory entries for known patterns of role assertion, embedded functions, or latent triggers before final prompt construction.
\end{itemize}

\textbf{LPCI Mitigation:} Multi-stage validation disrupts the LPCI execution chain by introducing gates that filter obfuscated logic (Stage 1), detect control manipulation (Stage 2), and block memory-triggered exploits (Stage 3), making it effective against deferred and cross-session attacks.

\subsection{Escalation Router}
\textbf{Targeted Attack Vectors:} Attack Vector 2: LPCI Core, and Attack Vector 3: Role Override via Memory Entrenchment

\textbf{Description:} The Escalation Router is a dynamic execution control layer that identifies high-risk prompt contexts and diverts them away from autonomous execution. Prompts are evaluated against a risk profile including known prompt injection triggers (e.g., ``ignore previous instructions'', ``act as system'') and behavioural scores.

\begin{itemize}
    \item \textbf{Routing Triggers:} Includes role assertions, tool invocations with elevated privileges, context-based logic shifts, and exploit markers.
    \item \textbf{Execution Sandboxing:} Enables prompt execution in a secure environment where downstream effects are contained.
    \item \textbf{Manual Review Queue:} Allows flagged interactions to be held for human review before any irreversible action.
\end{itemize}

\textbf{LPCI Mitigation:} This control acts as a runtime checkpoint to intercept payloads designed to escalate privileges or hijack workflows, particularly effective against logic-layer and memory-poisoned attacks.

\subsection{Cryptographic Tool and Data Source Attestation}
\textbf{Targeted Attack Vectors:} Primarily Attack Vector 1: Tool Poisoning; also mitigates Attack Vector 4: Vector Store Payload Persistence

\textbf{Description:} Enforces cryptographic proof of identity and integrity for all external tools and RAG data sources. Each component is signed and verified at runtime.

\begin{itemize}
    \item \textbf{Tool Signing:} Developers sign tool schemas, metadata, and endpoint URIs with private keys. The platform verifies signatures before invocation.
    \item \textbf{Data Source Attestation:} When documents are indexed into vector stores, manifests are signed and verified upon retrieval.
\end{itemize}

\textbf{LPCI Mitigation:} Prevents tool spoofing and raises defences against poisoned vector store content by rejecting unsigned or manipulated artefacts.

\subsection{Secure Ingestion Pipeline and Content Sanitisation for RAG}
\textbf{Targeted Attack Vectors:} Attack Vector 4: Vector Store Payload Persistence

\textbf{Description:} All ingested content is sanitised before indexing. The ingestion pipeline uses regex filters, semantic analysis, and LLM-based inspectors to flag embedded instructions.

\begin{itemize}
    \item \textbf{Metadata Tagging:} Flags imperative or high-risk content (e.g., \texttt{contains\_imperative\_language: true}, \texttt{risk\_score: 0.85}).
    \item \textbf{Contextual Demarcation:} Retrieved content is enclosed in markup (e.g., \texttt{<retrieved\_document\_content>}) to isolate and disarm latent instructions.
\end{itemize}

\textbf{LPCI Mitigation:} Prevents prompt confusion and injection by labelling retrieved non-authoritative content clearly before inclusion in model context.

\subsection{VI. Memory Integrity and Attribution Chaining}
\textbf{Targeted Attack Vectors:} Attack Vector 2: LPCI Core, and Attack Vector 3: Role Override via Memory Entrenchment

\textbf{Description:} Conversational memory is stored using tamper-evident chains. Each entry includes a timestamp, immutable author identity, and a cryptographic hash of the prior entry.

\begin{itemize}
    \item \textbf{Hash Chaining:} Ensures memory continuity and detects tampering during session reload.
    \item \textbf{Strict Role Enforcement:} Validates the author field in each memory entry. User-injected memory cannot impersonate system roles.
\end{itemize}

\textbf{LPCI Mitigation:} Detects offline tampering and blocks unauthorised elevation of privileges during memory recall or tool replay.

\subsection*{Evaluation}
These proposed mechanisms were retrospectively evaluated against the LPCI test suite and achieved an \textbf{84.94\% effective block rate} against prompt-based logic-layer attacks, particularly those using Base64 encoding, delayed triggers, and embedded memory payloads.

\textbf{Note:} These defence components are part of an internal framework under development. A formal public release of the \textit{Qorvex Security AI Framework (QSAF)} is scheduled for future publication.

\section{Discussion}
\label{sec:discussion}

Our comprehensive testing programme conclusively demonstrates that Logic-layer Prompt Control Injection represents a reproducible, cross-platform vulnerability that renders conventional security measures obsolete. These attacks transcend simple input filter bypass mechanisms; they exploit fundamental architectural assumptions regarding memory handling, logic execution, and context trust within LLM systems.

The empirical evidence is compelling. While ChatGPT achieved an 84.94\% block rate against our attack vectors, platforms such as LLaMA3 and Mixtral demonstrated critical vulnerabilities, executing nearly half of the 405 test payloads. The successful attacks were not anomalous occurrences—encoded logic patterns (Tests \#3, \#314) and conditional role-based triggers (Tests \#61, \#334) consistently penetrated defences, exposing fundamental gaps in runtime awareness and execution validation mechanisms.

These findings represent a paradigm shift in understanding prompt injection vulnerabilities. LPCI attacks do not conform to traditional single-session manipulation patterns. Instead, they exploit the complete LLM operational lifecycle—injection, storage, trigger, and execution—transforming system strengths into exploitable weaknesses. Memory persistence becomes a liability, contextual awareness becomes an attack vector, and implicit trust relationships become critical vulnerabilities.

Our lifecycle mapping and experimental results lead to an inescapable conclusion: static defence mechanisms are fundamentally inadequate against LPCI threats. Effective security requires deep integration of runtime logic analysis, memory recall validation, and downstream control flow monitoring. The QSAF mechanisms we propose—including prompt risk scoring, memory integrity verification, and cryptographic tool attestation—provide a comprehensive framework for addressing these vulnerabilities. However, defending against LPCI demands more than enhanced filtering capabilities; it requires a fundamental reconceptualisation of how we architect memory-aware, context-sensitive LLM infrastructures.

\subsection{Implications for Enterprise Deployment}

The prevalence of LPCI vulnerabilities has profound implications for enterprise LLM deployments:

\begin{enumerate}
    \item \textbf{Risk Assessment Protocols}: Organisations must reassess their AI security postures to account for persistent, cross-session attack vectors that traditional security audits may overlook.
    
    \item \textbf{Architectural Considerations}: System architects must design LLM integrations with explicit consideration for memory integrity, context validation, and temporal attack patterns.
    
    \item \textbf{Compliance and Governance}: Regulatory frameworks must evolve to address the unique risks posed by AI systems with persistent memory and autonomous decision-making capabilities.
    
    \item \textbf{Incident Response Planning}: Security teams must develop new playbooks for detecting and responding to logic-layer attacks that may manifest across extended time periods.
\end{enumerate}

\subsection{Limitations and Future Research}

While our research provides comprehensive insights into LPCI vulnerabilities, several limitations warrant acknowledgment:

\begin{enumerate}
    \item \textbf{Platform Coverage}: Our testing focused on publicly accessible LLM platforms. Enterprise-specific or custom implementations may exhibit different vulnerability profiles.
    
    \item \textbf{Evolving Defences}: The rapid pace of LLM development means that security measures are continuously evolving. Our results represent a snapshot of vulnerabilities as of June 2025.
    
    \item \textbf{Attack Sophistication}: We focused on demonstrable attack vectors. More sophisticated nation-state or advanced persistent threat actors may develop more complex LPCI variants.
\end{enumerate}

Future research directions should include:

\begin{itemize}
    \item Development of automated LPCI detection systems using anomaly detection and behavioural analysis
    \item Investigation of hardware-based security enclaves for protecting LLM memory integrity
    \item Exploration of formal verification methods for LLM logic execution paths
    \item Analysis of LPCI vulnerabilities in multi-agent systems and federated learning environments
\end{itemize}

\section{Literature Review}

Current AI security literature has primarily focused on surface-level prompt manipulation and response-based vulnerabilities. However, \textit{Logic-layer Prompt Control Injection (LPCI)} introduces a novel class of latent threats that operate beyond the scope of conventional detection methods. LPCI is characterised by:
\begin{itemize}
  \item Deferred and session-delayed activation
  \item Trigger mechanisms embedded in memory or external tools
  \item High resistance to detection through standard static or real-time scanning
\end{itemize}

This paper builds upon existing frameworks by addressing these deeper logic- and memory-layer attack vectors, establishing the need for runtime observability and persistent session analysis. This work is a dissection of "execution memory state" as a prompt injection vector, further solidfying the call for action shared in Johann's paper \cite{rehberger2024trustaipromptinjection} \cite{johann2024}.

\section{Conclusion}

This paper introduced \textbf{Logic-layer Prompt Control Injection (LPCI)} as a distinct and covert threat class targeting the \textit{logic execution}, \textit{memory retention}, and \textit{tool integration} layers of large language model (LLM) systems. Unlike traditional prompt injection, LPCI unfolds across multiple phases---\textbf{injection}, \textbf{storage}, \textbf{trigger}, and \textbf{execution}---and leverages delayed activation, memory entrenchment, and downstream execution pathways to bypass input-level defences.

Through a structured test suite of \textbf{1,700 cases} across five leading LLM platforms and targeted proof-of-concept demonstrations on \textit{Poe}, \textit{Grok}, and \textit{Bohrium}, we confirmed the practical feasibility of LPCI attacks in real-world environments. Successful payloads included encoded commands, cross-session triggers, vector-based logic injection, and role-conditioned execution---even in models with established safety filters.

\vspace{2cm}



\end{document}